# Role of electrodes in study of hydrovoltaic effects

Chunxiao Zheng[1], Sunmiao Fang[1], Weicun Chu[1], Jin Tan[1], Bingkun Tian[1], Xiaofeng Jiang[1], Wanlin Guo[1,2]*

[1]Key Laboratory for Intelligent Nano Materials and Devices of Ministry of Education, State Key Laboratory of Mechanics and Control of Mechanical Structures, Nanjing University of Aeronautics and Astronautics, Nanjing 210016, China

[2]Institute for Frontier Science of Nanjing University of Aeronautics and Astronautics, Nanjing 210016, China

*Corresponding author. Email: wlguo@nuaa.edu.cn

**Abstract**

The last decade has witnessed the emergence of hydrovoltaic technology, which can harvest electricity from different forms of water movement, such as raindrops, waves, flows, moisture, and natural evaporation. In particular, the evaporation-induced hydrovoltaic effect received great attention since its discovery in 2017 due to its negative heat emission property. Nevertheless, the influence of electrode reactions in evaporation-induced power generation is not negligible due to the chemical reaction between active metal electrodes and water, which leads to " exceptional " power generation. Herein, we designed a series of experiments based on air-laid paper devices with electrodes of different activities as the top and bottom electrodes. To verify the contribution of electrodes, we compared the output performance of different electrode combinations when the device is partially-wetted and fully-wetted. The device hydrophilicity, salt concentration, and acidity or basicity of solutions are also comprehensively investigated. It is demonstrated that the chemical reaction of active metals (Zn, Cu, Ag, etc.) with different aqueous solutions can generate considerable electrical energy and significantly distort the device performance, especially for Zn electrodes with an output voltage from ~1.26 to ~1.52 V and current from ~1.24 to ~75.69 μA. To promote the long-term development of hydrovoltaic technology, we recommend use of inert electrodes in hydrovoltaic studies, such as Au and Pt, especially in water and moisture environment.

**Keywords**：hydrovoltaic effects; electrodes; water; electrochemical reaction.

**Introduction**

The development of renewable green energy has become essential due to the global energy crisis and environmental pollution.[1] As the largest reserve of renewable energy on Earth, water exists in the form of vapor, liquid, and solid, covering about 71% of the Earth's surface and capturing about 70% of the solar energy arriving at the Earth's surface. Efficient collection and utilization of water resources could satisfy the global energy requirement.[2,3] In 2017, Xue et al. systematically demonstrated that water evaporation can consistently generate a sustained voltage of up to 1 V and a current of 100 nA from centimeter-sized carbon black films under ambient conditions.[4] Subsequently, the evaporation-induced hydrovoltaic effect has been intensively studied. Although both the water-movement-induced streaming potential as well as the direct interaction between the materials and evaporating water molecules plays a role,[4,5] the underlying mechanism is still being widely investigated to enhance the output performance. However, the chemical reaction between the active metal electrodes and water may lead to false exceptional high hydrovoltaic effects.[6-8]

It is important to distinguish whether the electricity is generated by evaporation-induced hydrovoltaic effect or by chemical reactions between electrodes and water. The evaporation-induced hydrovoltaic effect directly converts latent heat into electrical energy spontaneously and continuously.[4,9] However, the redox reaction of the active electrode with water may act as one kind of metal-air battery, which irreversibly converts electrochemical energy into electricity. The metal-air battery comprises a metal anode, an air cathode, and a separator soaked in metal-ion conducting electrolytes.[10] The metal anode is oxidized and releases electrons to the external circuit when the cell is discharged. At the same time, oxygen diffuses into the cathode, accepts the electrons from the anode, and is reduced to oxygen-containing species. This kind of metal-air battery is disposable with limited lifespans and cannot maintain stable power generation.[11] Therefore, the use of active metal electrodes might lead to confusion between the two concepts. For example, for a flexible dual-mode electricity nanogenerator based on the synergy of streaming potential and iron-air primary battery, the majority of the energy obtained from aqueous solution comes from the electrochemical reaction of iron electrode in the solution.[11] When carbon black (CB) ink-modified a wood sponge immersed in LiCl solution was measured with copper electrodes, the measured output power of 216 μW contains the main contribution of the copper-moisture primary battery.[12] The reported carbon/metal based water-evaporation-generator featuring carbon (top) and metal (bottom) electrodes also presented a discussion related to the corrosion of metal electrodes, demonstrating that metal corrosion empowered water-evaporation-generator.[13] Metal corrosion inevitably leads to considerable power generation, mixing with the evaporation-induced hydrovoltaic effect based on the electrical double layer (EDL) or evaporating potential. Therefore, a good understanding of the role of electrodes is essential for developing hydrovoltaic devices.

To verify the contribution of chemical reactions of active metals with water or air to the output performance, we systematically investigated the power generation of typical two-end hydrovoltaic devices composed of air-laid paper (AP) coated with porous CB and polyvinyl alcohol (PVA) using different activities electrodes as the top and bottom electrodes. This AP device with inert electrodes itself has low output within 30 mV @ 10 nA, so that the contribution of active electrodes can be easily distinguished. We compared the output performance of different combinations of electrode materials with the device in partially-wetted and fully-wetted states. The influences of hydrophilicity, salt concentration, and acidity or basicity of solutions on power generation are comprehensively investigated. It is shown that the proper use of electrodes is essential for the healthy development of hydrovoltaic technology.

**Results**

The fabrication process of the PVA/CB/AP membrane is illustrated in Figure 1a. The AP (~ 200 μm in thickness) substrate woven by cellulose fiber is dip-coated in CB suspension and PVA solutions successively, where CB as a classical evaporation material has been extensively studied and the bonding effect of PVA between the interfaces prevents the shedding of CB.[14] Scanning electron microscopy (SEM) images show that homogenous cellulose fibers form stable three-dimensional interconnected nanochannels with CB and PVA being tightly attached (Figure 1b-d). Figure 1e illustrates the typical experimental setup of the evaporation-induced device, in which two electrodes are attached to the bottom (bottom electrode) and top (top electrode) ends of the PVA/CB/AP sheet of size 1×10 cm$^2$. 100 µl of DI water was added to the bottom end of the membrane maintaining it in partially-wetted state, i.e., the bottom electrode is wet, and the top electrode remains dry in the air (Figure 1e) during the following tests without a specific statement. To investigate the influence of electrode reaction on power generation, a series of conductive materials, such as aurum (Au), platinum (Pt), carbon nanotubes (CNTs), argentum (Ag), copper (Cu), and zinc (Zn), were used as the top or bottom electrode of the device to measure the open-circuit voltage ($V_{OC}$) and short-circuit current ($I_{SC}$) as shown in Figures 1f-g. The output performance of the device using inert electrodes is negligibly low (Au ~ 25 mV, 12 nA, Pt ~ 30 mV, 47 nA) in comparison with the $V_{OC}$ and $I_{SC}$ values using Zn, Cu, and Ag electrodes which are up to 1.3 V and 1.5 µA, 0.33 V and 0.25 µA, and 0.18 V and 0.19 µA, respectively. Such dramatic increases in $V_{OC}$ and $I_{SC}$ of devices with Zn, Cu, and Ag bottom electrodes are mainly ascribed to the electrochemical reactions of the active metals at the wet interface. Therefore, using the active metal electrodes at the wet bottom end, the intrinsic weak hydrovoltaic effect can be falsely exaggerated. Even the commercial CNTs electrode can also affect the output performance obviously depending on the impurities (Figures 1f and 1g).

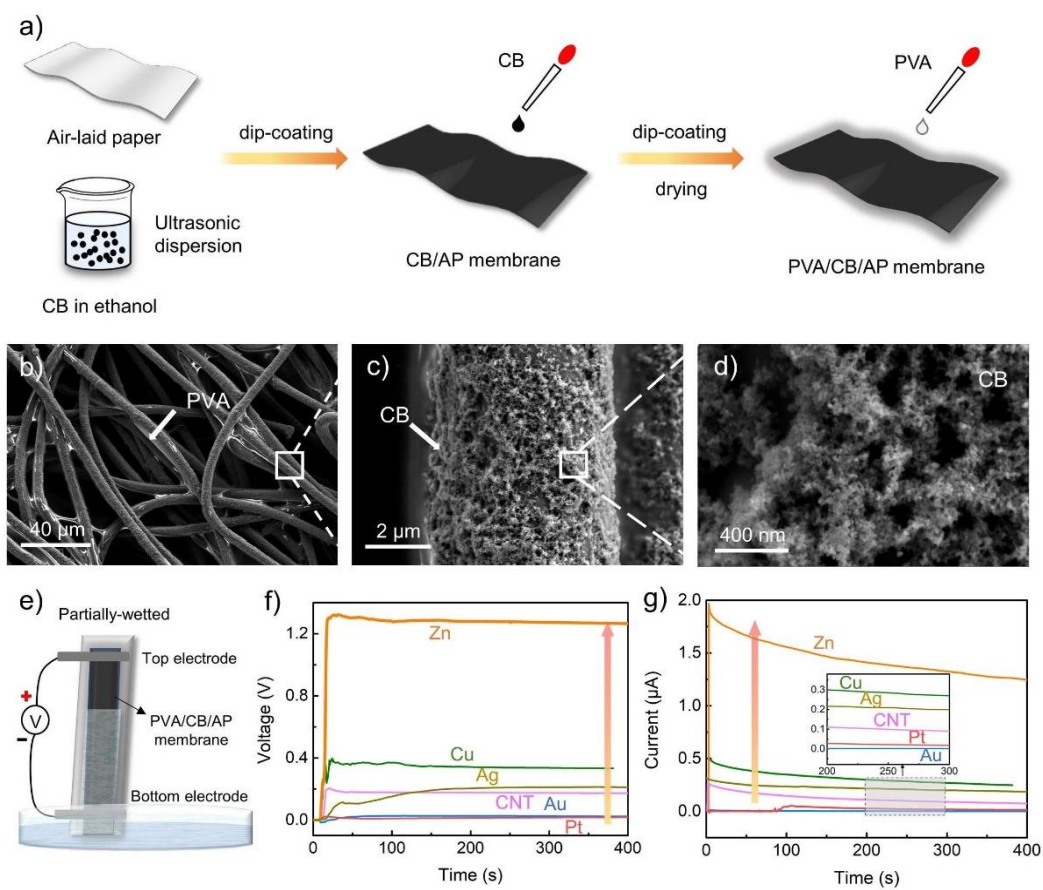

Figure 1. Manufacture of devices and the role of different electrodes in DI water. (a) The preparation process of the PVA/CB/AP membrane; (b-d) SEM images of the PVA/CB/AP membrane; (e-g) Schematic diagram of the two-electrode PVA/CB/AP device (e) and corresponding voltage (f) and current values (g) with different electrode materials. The top (in air) and bottom electrodes (in DI water) use the same materials in these experiments. (RH=45%)

The reactions of a typical device with the top Au electrode in air and the bottom Zn electrode wet by DI water (Au-Zn) are shown in Figure 2a. The Raman spectra of the Zn electrode before and after measurement (has been wetted) are shown in Figure 2b. The pristine Zn electrode lacked characteristic peaks, while the characteristic peak of ZnO centered at 99 $cm^{-1}$ occurred after 12 h of oxidation. The appearance of the peak centered at 586 $cm^{-1}$ was ascribed to the intrinsic defects of ZnO (oxygen vacancies, Zn interstitial), indicating that an obvious redox reaction has occurred between the Zn electrode and water.[15]

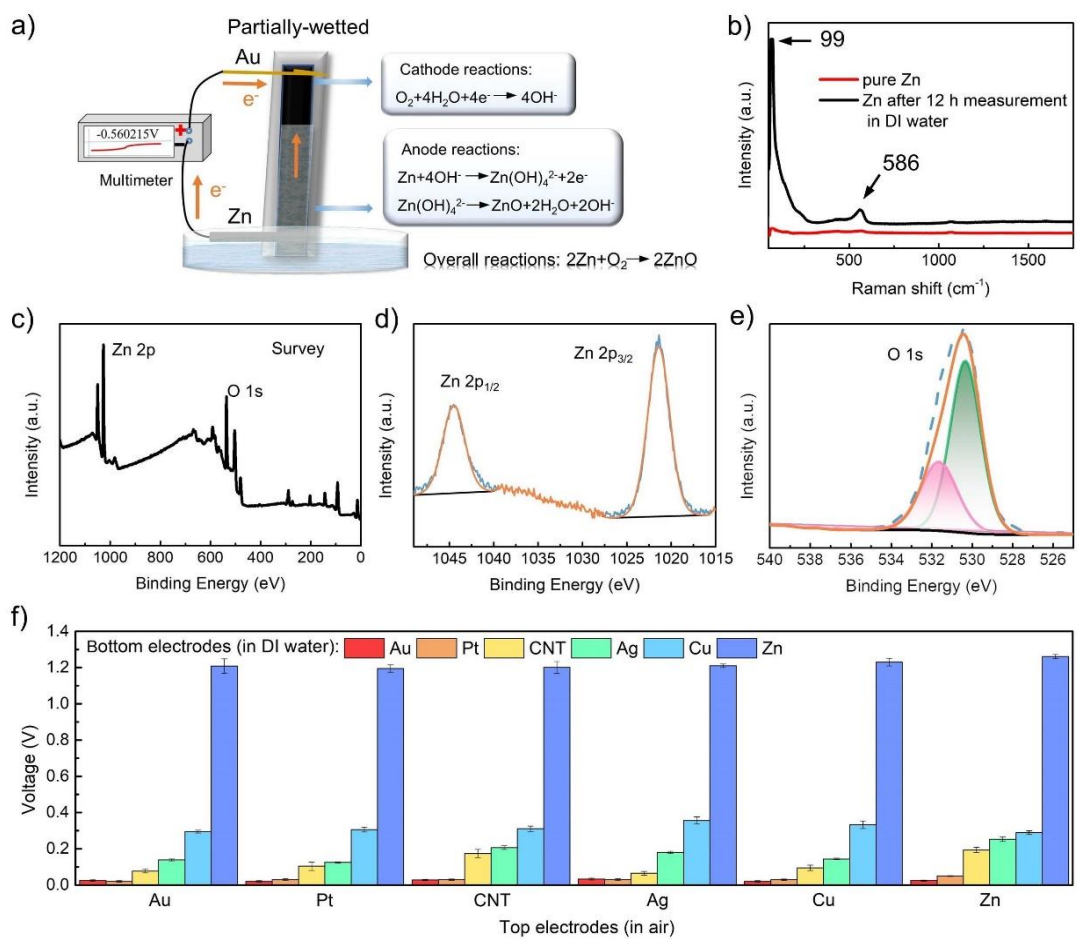

Figure 2. Two-electrode devices with the top one in ambient air and the bottom one in DI water. (a) Demonstration of origin of power generation with Au and Zn as the top and bottom electrodes respectively (Au-Zn); (b) Raman spectra of the Zn electrode before and after electricity measurement; (c-e) XPS spectra of the Zn electrode after oxidation: (c) Survey scan, (d) Zn 2p peaks, (e) O 1s peak; (f) The $V_{OC}$ and $I_{SC}$ of the PVA/CB/AP membrane with the top electrode in air and the bottom electrode in DI water. Different materials are used as the top and bottom electrodes during measurement.

The oxidation of the Zn electrode is further confirmed by SEM and Energy Dispersive X-Ray Spectroscopy (EDX) as shown in Figures S1 and S2. The SEM images of the oxidated Zn show a rough surface in comparison to that of the pristine Zn electrode. The O element of the oxidated Zn marked a significant increase after oxidation as revealed by EDX characterizations. In addition, X-ray photoelectron spectroscopy (XPS) spectra show that the Zn electrode reacted with water in a redox reaction. Figure 2c exhibits the XPS spectrum of the Zn electrode after oxidation that manifests the existence of the element Zn and O. The two strong peaks at 1021.4 and 1044.4 eV belong to the Zn $2p^{3/2}$ and Zn $2p^{1/2}$ states, respectively (Figure 2d). The O 1s spectrum in Figure 2e is deconvoluted into two peaks at 530.3 and 531.3 eV, which is attributed to the oxygen atoms coordinated with Zn atoms and hydroxyl oxygen on the sample surface, respectively.[16]

We further investigated the output performance of different combinations of the six electrodes. The $V_{OC}$ and $I_{SC}$ values of the partially-wetted PVA/CB/AP membrane with different electrode combinations are shown in Figure 2f. It can be observed that the measured output of the device with Zn electrode is significantly higher than other electrodes. In comparison, the measured output power with Au and Pt electrodes is extremely low, indicating that the influence of chemical reactions on the hydrovoltaic effect can be suppressed efficiently by use of inert Au and Pt electrodes for several orders of magnitude to replace the active Ag/Cu/Zn electrodes, especially the most active Zn electrode. The output of the redox reaction of active electrodes with water is several orders higher than the real evaporation-induced power generation. Therefore, in developing hydrovoltaic devices, the influence of active electrodes should be carefully excluded by using inert materials and the contribution of possible metal-air batteries should be distinguished from the evaporation-induced hydrovoltaic effect. The reactions of metal-air batteries have no relationship with the mechanisms of hydrovoltaic effects. Active electrodes were blindly used to improve the output performance of hydrovoltaic devices, or the electrochemical reactions were used to explain the mechanism of hydrovoltaic effects, which confused the two concepts and is misleading.

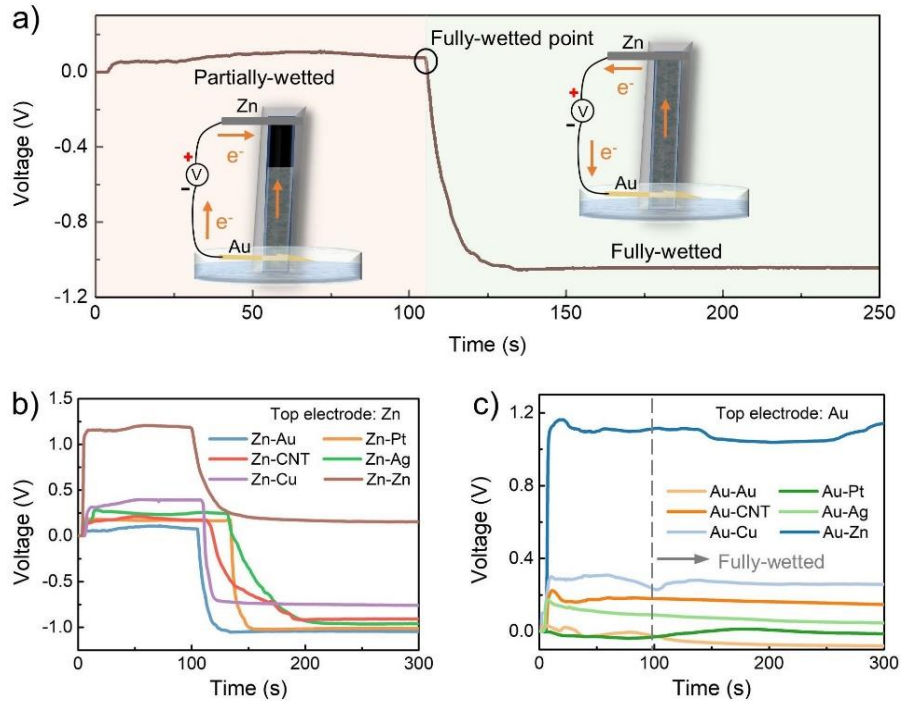

Figure 3. Two-electrode devices with the bottom electrode initially in and both electrodes then immersed into DI water. (a) Typical voltage variation when the PVA/CB/AP device shifts from partially-wetted state to fully-wetted state with Au and Zn as the bottom and top electrodes respectively. The dashed line indicates the time when the device is fully wetted; (b-c) Voltage variations of the PVA/CB/AP device with Zn (b) as the bottom electrode and Au (c) as the top electrode when the device shifts from partially-wetted state to fully-wetted state. (~100 s in the fully-wetted state)

In fully-wetted states, it is further intuitive to observe the significant impact of the redox reaction of the active electrode with water on the evaporation-induced power generation in Figures 3 and S3. Firstly, if the top electrode was an active metal electrode and the bottom electrode was inert, e.g. the top electrode was Zn electrode and the bottom electrode was Au electrode, the measured $V_{OC}$ was 0.1 V in a partially-wetted state, which could be attributed to the classical streaming potential in evaporation-induced hydrovoltaic effect. Interestingly, the $V_{OC}$ shifted from 0.1 V to -1.1 V when the top electrode (Zn) contacts the water, and a fully-wetted turning point can be distinctly observed. The directional change and increase of the electrical signal indicated that the redox reaction between the Zn electrode and water dominated power generation with the formation of the metal-air battery at this point, which covers the weak evaporation-induced hydrovoltaic effect.

For the Zn-Zn electrode configuration, the measured *V*oc is about 1.2 V in partially-wetted state (Figure 3b), which is a typical metal-air battery. The *V*oc immediately decreases from 1.2 V to 0 when the top electrode contacts water, in which the asymmetric structure of the metal-air battery is broken. The Zn electrode is the most active among the six electrodes. Therefore, for the Zn-other electrode configuration, the polarity of the voltage changes obviously once the top electrode (Zn) contacts water, which can be ascribed to a reversed Zn-air chemical reaction. In contrast, when inert Au is used as both the top and bottom electrodes (Figure 3c), no significant voltage could be generated, indicating that the evaporation-induced hydrovoltaic effect as well as the chemical reaction is quite weak in this system. With Au as the top electrode and an active metal as the bottom electrode, a metal-air battery is formed immediately when the bottom electrode contacts water. Subsequently, the value of $V_{OC}$ remains stable in fully-wetted state where no evaporation is involved. Furthermore, the I-V curves comparison of Au and Zn electrodes in the PVA/CB/AP device further verifies the above experimental phenomena (Figure S4), in which the redox reactions play a dominant role. Therefore, it is essential to exclude any possible chemical reactions in study of the hydrovoltaic effect.

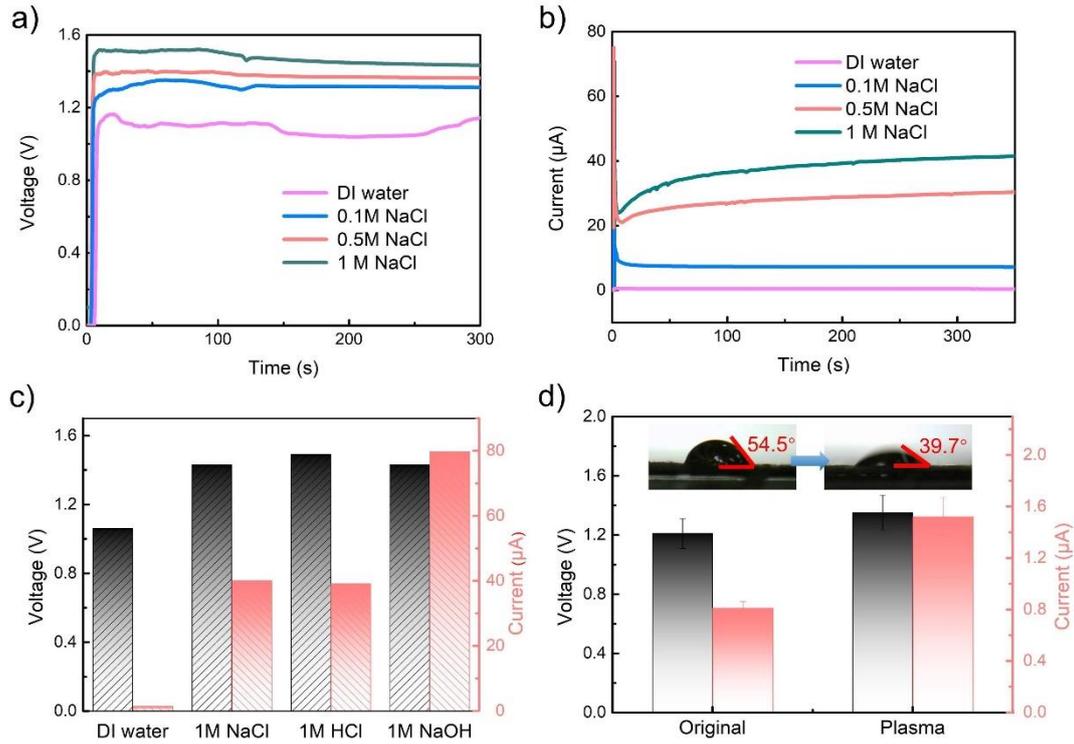

Figure 4. Devices with Zn as the bottom electrode in different aqueous solutions and Au as the top electrode in ambient air. (a-b) The $V_{oc}$ (a) and $I_{sc}$ (b) of the device with solutions of different NaCl concentrations; (c) The $V_{oc}$ and $I_{sc}$ of the device in different solutions; (d) The $V_{OC}$ of the device before and after plasma treatment.

The surface chemistry, salt concentration, and acidity or basicity of solutions may have a significant impact on the redox reaction of the electrodes. It is well known that metal-air batteries usually converted chemical energy into electric energy in saline solutions. So we further investigated the output performance of the PVA/CB/AP device with Au-Zn electrodes in different saline solutions. In Figures 4a-b, under a range of NaCl concentrations (from 0.1 to 1 M), the $V_{OC}$ and $I_{SC}$ increase with increasing concentration due to faster charge transfer in the NaCl solution with higher ion concentration.[11] In addition, the $Cl^-$ in the NaCl solution could accelerate the oxidation reaction of Zn to generate more $Zn^{2+}$, reducing the resistance between the Au and Zn electrodes.[12] As a result, increasing the saline concentration of the electrolyte enhances the measured output, which is inconsistent with the law of evaporation-induced electricity generation.[4] The oxidation of the Zn electrode after measurement in the NaCl solution can also be confirmed by the SEM images shown in Figure S5. In addition, Cu electrodes are commonly used in hydrovoltaic devices. As shown in Figure S6, the oxidation of the Cu electrode is confirmed by SEM and EDX similar to the case of the Zn electrode.

We further investigated the output of the device in different solutions (DI water, 1 M NaCl, 1 M NaOH, and 1 M HCl), as shown in Figure 4c. The output of the device in 1 M HCl and NaCl solution is similar but higher than that in DI water, mainly because the Cl$^-$ of the same concentration facilitated the redox reactions of the Zn electrode. The key difference is that the oxidation of the Zn electrode occurs in saline solutions ($2Zn+O_2 \rightarrow 2ZnO$), while hydrogen evolution occurs in acidic solutions, i.e., $Zn+2HCl \rightarrow ZnCl_2+H_2$. The $I_{SC}$ of the device in 1 M NaOH solution increases significantly (~80 µA), as NaOH electrolyte could promote the dissociation of oxygen-containing groups into negatively charged groups, inducing a more negative surface potential and accelerating the rate of the electron transport for the redox reaction. The $V_{oc}$, $I_{sc}$, and resistance values for different electrode configurations in different solutions are compared in Table S1. Therefore, the combined effects of the activity of electrodes and the electrolytes should be considered in developing hydrovoltaic devices.

The PVA/CB/AP membrane was subsequently treated by $O_2$ plasma for 1 min at 25 W, introducing oxygen-containing functional groups to the surface. The as-treated PVA/CB/AP membrane exhibited enhanced hydrophilicity, as demonstrated by a reduction in the water contact angle from 54.5° to 39.7°, which could be attributed to the successful introduction of abundant oxygen-containing groups on the surface. As shown in Figure 4d, the PVA/CB/AP device after $O_2$ plasma treatment with Au (top) and Zn (bottom) electrodes generates remarkably higher $V_{OC}$~1.35 V and $I_{SC}$~1.5 µA, demonstrating that oxygen-containing functional groups facilitate the redox reaction and enhance the output.

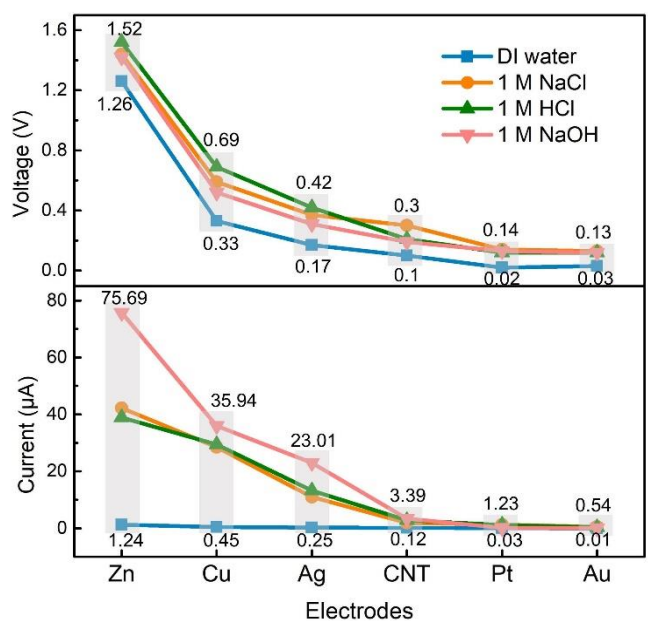

Figure 5. The measured output of the devices with different electrodes in different aqueous solutions. The top (in air) and bottom electrodes (in aqueous solutions) are the same material in these experiments.

To compare the specific output ranges provided by the six electrodes in the device, we summarized the measured output of different electrodes in four different aqueous solutions (DI water, 1 M NaCl, 1 M NaOH, and 1 M HCl). As shown in Figure 5, the output voltage of the Zn electrode ranges from ~1.26 to ~1.52 V, and the output current ranges from ~1.24 to ~75.69 μA, indicating that the Zn electrode reacts strongly with four aqueous solutions. Cu and Ag electrodes are less active than the Zn electrode, and their output ranges decrease from ~0.33 to ~0.69 V, ~0.45 to ~35.94 μA and ~0.17 to ~0.42 V, ~0.25 to ~23.01 μA, respectively. In addition to active metals, the commercial CNT electrode with aqueous solutions could also affect the output to some extent due to impurities generated by the production process (~0.1 to ~0.3 V and ~0.12 to ~3.39 μA). XPS analysis of the CNT electrode shows signals for C and O (Figure S7). The atomic percentage of O is 17.2%. The C1s peak can be decomposed into four components. In detail, the peak centered at 286.08, 287.38, and 288.68 eV corresponds to C-O, C=O, and -COO-, respectively. Therefore, if the active metal, commercial carbon, or other electrodes with similar activity were used, the measured output of the device should be much higher than its intrinsic hydrovoltaic effects without the contribution of chemical reactions. For developing the evaporation-induced hydrovoltaic devices, the Au and Pt, or highly purified carbon electrodes are recommended, whose electrode reactions with the aqueous solution can be negligible weak.

**Conclusion**

In summary, we have systematically investigated the output performance of carefully designed hydrovoltaic devices with electrodes of different activity combinations in the partially-wetted and fully-wetted states. The influences of hydrophilicity, saline concentration, and acidity or basicity of solutions are also investigated. The background output of the device with inert electrodes is 30 mV@10 nA. By comparing the measured output as well as the polarity of the electrical signal, it is found that the electrochemical reactions of active metals (Zn, Cu, Ag, etc.) with water or aqueous solutions can generate considerable electrical energy and significantly distort the hydrovoltaic performance of the device. Especially when Zn electrode is used, the distortion in the output voltage can be up to ~1.52 V and the distortion in the output current can be up to ~76.69 μA. Under this circumstance, the device behaves as a typical metal-air battery, rather than a hydrovoltaic device. Even the commercial carbon electrodes in aqueous solutions can also distort the output performance up to about 0.1 V@0.12 μA in this study. Therefore, use of inert electrodes, such as Au and Pt, or highly purified carbon ones, is recommended for the study of hydrovoltaic effects. Excluding the distortion of the chemical reaction of the electrodes to the electricity generation performance can facilitate the long-term and healthy development of hydrovoltaic technology.

**Materials**

Air-laid paper (AP) was obtained with Suzhou Ketong Purification Technology Co., Ltd., China. Conductive carbon black particles were provided by Nanjing XFNANO Materials Tech Co., Ltd. Polyvinyl alcohol (PVA, Mw = 67000 g/mol) was purchased from Aladdin Industrial Co.

**Fabrication of the PVA/CB/AP membrane**

A simple dip-coating method is used to prepare the PVA/CB/AP membrane. Firstly, AP was ultrasonically cleaned in ethanol for 20 min and then oven-dried at 80°C. Subsequently, 100 mg of CBs with an average diameter of 30-45 nm and a specific surface area of 100-110 $m^2/g$ were dispersed in 200 mL of ethanol and formed into a homogeneous suspension after 1.5 h ultrasonication at 300 W. Afterward, the prepared carbon black suspension was applied onto the AP by dip-coating repeatedly to form a CB/AP membrane. Finally, dip-coating and oven-dried PVA/CB/AP membrane was assembled with an aqueous 2wt% PVA solution onto the CB/AP membrane.

**Characterization**

The morphologies of the samples were tested by Field Emission Scanning Electron Microscopy (FE-SEM, JSM-7600F, JEOL, Japan). X-ray photoelectron spectroscopy (XPS) was performed using an Escalab250Xi photoelectron spectrometer with Al Kα (1486.6 eV). Energy-dispersive X-ray spectra were performed using an Octane Elect EDX system. A Keithley DMM6500 multimeter measured the real-time open-circuit voltage and short-circuit current. The I-V curves using a CHI760E electrochemical workstation (Chenhua, China) were performed on a PVA/CB/AP membrane with a 1x10 $cm^2$.

**Acknowledgments**

This work was supported by the National and Jiangsu Province NSF (T2293691, BK20212008) of China, National Key Research and Development Program of China (2019YFA0705400), the Research Fund of State Key Laboratory of Mechanics and Control of Mechanical Structures (MCMS-I-0422K01), the Fundamental Research Funds for the Central Universities (NJ2022002) and the Fund of Prospective Layout of Scientific Research for NUAA (Nanjing University of Aeronautics and Astronautics).